\documentclass[10pt,twocolumn]{IEEEtran}

\usepackage{cite}
\usepackage{graphicx}
\usepackage{epsfig}
\usepackage{psfrag}
\usepackage{subfigure}
\usepackage{url}
\usepackage{stfloats}
\usepackage{amsmath}
\usepackage{color}
\usepackage{amssymb}
\usepackage{algorithm}
\usepackage{algorithmic}
\usepackage{multirow}
\usepackage{array}
\usepackage{booktabs}
\usepackage{multirow}
\usepackage{threeparttable}

\begin{document}
%
\title{Photoplethysmography-Based Heart Rate Monitoring in Physical Activities via Joint Sparse Spectrum Reconstruction}

\author{Zhilin Zhang, \IEEEmembership{Senior Member, IEEE}
\thanks{Z. Zhang is with the Emerging Technology Lab, Samsung Research America -- Dallas, 1301 East Lookout Drive, Richardson, TX 75082, USA. Email: zhilinzhang@ieee.org. }
\thanks{Copyright (c) 2015 IEEE. Personal use of this material is permitted. However, permission to use this material for any other purposes must be obtained from the IEEE by sending an email to pubs-permissions@ieee.org.}
}

\markboth{Published in IEEE Transactions on Biomedical Engineering, Vol. 62, No. 8, PP. 1902-1910, August 2015}{Zhang \MakeLowercase{\textit{et al.}}: }

\maketitle

\begin{abstract}
\emph{Goal}: A new method for heart rate monitoring using photoplethysmography (PPG) during physical activities is proposed. \emph{Methods}: It jointly estimates spectra of PPG signals and simultaneous acceleration signals, utilizing the multiple measurement vector model in sparse signal recovery. Due to a common sparsity constraint on  spectral coefficients, the method can easily identify and remove spectral peaks of motion artifact (MA) in PPG spectra. Thus, it does not need any extra signal processing modular to remove MA as in some other algorithms. Furthermore, seeking spectral peaks associated with heart rate is simplified. \emph{Results}: Experimental results on 12 PPG datasets sampled at 25 Hz and recorded during subjects' fast running showed that it had high performance. The average absolute estimation error was 1.28 beat per minute and the standard deviation was 2.61 beat per minute. \emph{Conclusion and Significance}: These results show that the method has great potential to be used for PPG-based heart rate monitoring in wearable devices for fitness tracking and health monitoring.

\end{abstract}


%

\IEEEpeerreviewmaketitle

\section{Introduction}

With the emerging of smart-watches and smart wristbands for healthcare and fitness, heart rate (HR) monitoring using photoplethysmography (PPG) \cite{allen2007PPG,tamura2014PPGreview} recorded from wearers' wrists becomes a new research topic both in industry and academia. HR monitoring using PPG signals has many advantages over traditional ECG signals, such as simpler hardware implementation, lower cost, and no requirement of ground sensors and reference sensors as in ECG recordings.

However, during physical activities motion artifact (MA) contaminated in PPG signals seriously interferes with HR estimation. The MA is mainly caused by ambient light leaking into the gap between a PPG sensor surface and skin surface. The gap can be easily enlarged by hand movements during physical activities. Besides, the change in blood flow due to movements is another MA source \cite{maeda2011relationship}. Consequently, the PPG signal contains strong MA, while the heartbeat-related PPG component is weak. Therefore, estimating heart rate becomes very challenging. In \cite{TROIKA,Zhang2014GlobalSIP}, several typical difficulties in HR estimation are discussed.

So far, many noise-reduction techniques have been proposed, such as independent component analysis (ICA) \cite{kim2006ICA}, adaptive noise cancelation (ANC) \cite{poh2010motion,yousefi2013ANC}, spectrum subtraction \cite{fukushima2012SpectrumSubtraction}, Kalman filtering \cite{lee2010kalman}, wavelet denoising \cite{raghuram2010wavelet}, and empirical mode decomposition \cite{sun2012robust}, to name a few. But these techniques were mainly proposed for  scenarios when MA is not strong.

The TROIKA framework \cite{TROIKA} is a recently proposed method for scenarios when MA is extremely strong. It consists of signal decomposition, sparsity-based high-resolution spectrum estimation, and spectral peak tracking and verification. The signal decomposition aims to partially remove MA components which overlap with the heartbeat-related PPG components in the same frequency band,  and also to sparsify the PPG spectra facilitating the sparsity-based high-resolution spectrum estimation. The spectral peak tracking with verification aims to correctly select the spectral peaks corresponding to HR, overcoming various unexpected situations such as  nonexistence of HR-related spectral peaks. Experimental results have shown that the TROIKA framework has high performance during wearers' intensive physical activities.

In this paper  a new approach to HR estimation is proposed, which is based on JOint Sparse Spectrum reconstruction, denoted by JOSS. It exploits the fact that the spectra of PPG signals and simultaneous acceleration signals have some common spectrum structures, and thus formulates the spectrum estimation of these signals into a joint sparse signal recovery model, called the multiple measurement vector (MMV) model  \cite{Cotter2005}. Although the idea of using basic sparse signal recovery algorithms was initially proposed in  TROIKA \cite{TROIKA}, the present approach is novel and has many advantages over TROIKA. The main novelty is the use of the MMV model for joint spectrum estimation, in contrast to the single measurement vector (SMV) model \cite{TROIKA} in TROIKA. In the MMV model, the measurement vectors are PPG signals and acceleration signals. Thus, their spectra are estimated simultaneously. Using the MMV model has many benefits which cannot be obtained when using the SMV model:
\begin{itemize}
  \item Based on theoretical analysis, given the same sparsity level and compression ratio, the MMV model is known to have much better reconstruction performance than the SMV model \cite{Cotter2005,Jin2013MMV,Zhang2011IEEE}.

  \item Due to the common sparsity constraint of the MMV model \cite{Cotter2005} on the spectral coefficients, the spectral peaks of MA in the raw PPG spectra can be easily found by checking  spectral peaks in the acceleration signal spectra at  corresponding frequency bins.

  \item By comparing the spectral coefficients in the PPG spectrum and the acceleration signal spectra at the same frequency bins, we can cleanse the PPG spectra, achieving a similar result as spectral subtraction \cite{boll1979suppression}.
\end{itemize}
Due to the above benefits, the proposed approach does not require the signal decomposition and the temporal difference operations  in TROIKA. Furthermore, the spectral peak tracking and verification in TROIKA can be largely simplified. Thus, the approach is more suitable for hardware implementation \cite{Xu2013EL}.

The rest of the paper is organized as follows. Section \ref{sec:algorithm} presents the approach. Section \ref{sec:experiments} gives experimental results. Discussion and conclusion are given in the last two sections.

\section{Proposed Method}
\label{sec:algorithm}

This section  first presents the motivations, and then proposes the JOSS method. The JOSS method has two parts. One part is joint sparse spectrum reconstruction using the MMV model,  followed by a simple spectral subtraction. The second part is spectral peak tracking with verification. Note that before fed into JOSS, PPG signals and acceleration signals are first bandpass filtered.

\subsection{Motivations}
\label{subsec:motivation}

In \cite{TROIKA} the following SMV model was used to estimate the sparse spectrum of a raw PPG signal,
\begin{eqnarray}
\mathbf{y} = \mathbf{\Phi} \mathbf{x} + \mathbf{v},
\label{eq:SMV}
\end{eqnarray}
where $\mathbf{y} \in \mathbb{R}^{M \times 1}$ is a segment of a raw PPG signal,  $\mathbf{\Phi} \in \mathbb{C}^{M \times N} (M<N)$ is a redundant discrete Fourier transform (DFT) basis, $\mathbf{x} \in \mathbb{C}^{N \times 1}$ is the desired solution vector, and $\mathbf{v} \in \mathbb{R}^{M \times 1}$ models measurement noise or modeling errors. The redundant DFT basis is given by
\begin{eqnarray}
\Phi_{m,n} = e^{j \frac{2\pi}{N}mn}, \; m = 0,\cdots,M-1; \, n=0,\cdots,N-1
\label{eq:DFTbasis}
\end{eqnarray}
where $\Phi_{m,n}$ denotes the $(m,n)$th entry of $\mathbf{\Phi}$. In the SMV model, a key assumption is that $\mathbf{x}$ is sparse or compressive, i.e. most elements of $\mathbf{x}$  are zero or nearly zero, while only a few elements have large nonzero values.

Based on this model, the estimated $k$th spectrum coefficient of the PPG signal,  denoted by $s_k$, is thus given by
\begin{eqnarray}
s_k = | \widehat{x}_k |^2, \quad k=1,\cdots,N
\end{eqnarray}
where $\widehat{x}_k$ is the $k$th element of $\widehat{\mathbf{x}} \in \mathbb{C}^{N \times 1}$, the estimate of $\mathbf{x}$.

Using sparse signal recovery algorithms to estimate spectra has  advantages over conventional nonparametric spectrum estimation methods and line spectral estimation methods, such as high spectrum resolution, low estimation variance, and increased robustness \cite{Duarte2013spectral}.

In TROIKA \cite{TROIKA} a raw PPG was first bandpass filtered and then cleansed by partially removing MA via a signal decomposition procedure. Then the cleansed PPG signal was processed by a temporal difference operation to further suppress low-frequency MA and enhance the harmonic components of heartbeat-related PPG components. Finally, the resulting signal was used to calculate its spectrum by an SMV-model-based sparse signal recovery algorithm.

However, the signal decomposition procedure  cannot remove all prominent MA components. It has two drawbacks.

First, in this procedure, a spectral peak of MA in a raw PPG spectrum is identified by checking if there is also a spectral peak in an acceleration spectrum at the same frequency bin. However,  both the PPG spectrum and the acceleration spectrum are calculated  using Periodogram separately. Consequently, even caused by a common hand movement, the spectral peak in the PPG spectrum and the counterpart peak in the acceleration spectrum may not appear at the same frequency bin. (They may locate at two frequency bins that are close to each other.) Thus, the MA spectral peak in the raw PPG spectrum is not identified, and will remain in the PPG spectrum calculated by sparse signal recovery in a later stage.

Second, if an MA component has a dominant spectral peak locating close to the frequency bin at which the previously selected HR-related spectral peak locates, the MA component will be kept in the PPG signal.

To overcome the drawbacks, this work proposes using MMV-model-based sparse signal recovery to estimate  spectra of PPG and acceleration signals jointly.  Based on this method, one can easily and reliably cleanse  PPG spectra by removing MA spectral peaks, similar as spectral subtraction.

\subsection{Joint Sparse Spectrum Reconstruction Using the MMV Model}
\label{subsec:MMV}

The MMV model is an extension of the SMV model, estimating solution vectors jointly from multiple measurement vectors. In estimating power spectra of multichannel signals, the model can be expressed as follows
\begin{eqnarray}
\mathbf{Y} = \mathbf{\Phi} \mathbf{X} + \mathbf{V},
\label{eq:MMV}
\end{eqnarray}
where $\mathbf{Y} \in \mathbb{R}^{M \times L}$ is the matrix consisting of $L$ measurement vectors, $\mathbf{\Phi}$ is the redundant DFT basis as before, $\mathbf{X} \in \mathbb{C}^{N \times L}$ is the desired solution matrix, and $\mathbf{V} \in \mathbb{R}^{M \times L}$ represents measurement noise or model errors.
A key assumption in the MMV model is that the solution matrix $\mathbf{X}$ is row-wise sparse, that is, only a few rows in $\mathbf{X}$ are nonzero while most  rows are zero. This assumption is also referred to as the `common sparsity constraint' \cite{Cotter2005}.

To use the MMV model, the columns of the measurement matrix $\mathbf{Y}$ are segments of PPG and simultaneous acceleration signals in the same time window. In the experiments of this work, $\mathbf{Y}$ is formed by a segment of one channel of PPG and segments of three channels of simultaneous acceleration signals. Thus, $L=4$. Each column of $\mathbf{X}$ yields the spectrum of the corresponding signal.

Many sparse signal recovery algorithms have been proposed for this model \cite{Cotter2005,Zhang2011IEEE}. But not every algorithm is suitable, since here the matrix $\mathbf{\Phi}$ is highly coherent, \emph{i.e}., high correlation exists between neighboring columns of $\mathbf{\Phi}$. This work adopts the Regularized M-FOCUSS algorithm \cite{Cotter2005}, because it has fast speed and reliable performance even if $\mathbf{\Phi}$ is highly coherent.

The joint sparse signal recovery has many advantages. Using the MMV model, we can recover unique solutions which are less sparse. For example, in \cite{Cotter2005} it is shown that the MMV model can ensure a unique solution with the number of nonzero rows $r_0$ satisfies
      \begin{eqnarray}
        r_0 \leq \lceil (M+L)/2   \rceil - 1
      \end{eqnarray}
where $\lceil \cdot \rceil$ denotes the ceiling operation. Instead, the SMV model can only recover unique solutions with the nonzero coefficient number $r_0 \leq \lceil (M+1)/2   \rceil - 1$. These theoretical results are inspiring. In the present application problem,  the number of spectral coefficients sometimes is large due to MA. Although bandpass filtering is generally used before spectrum estimation, the spectral coefficients in the pass band are still many, which is an adverse situation. The TROIKA framework uses signal decomposition to remove MA components in the pass band to sparsify the spectrum, but this procedure is not always effective as stated in Section \ref{subsec:motivation}. In contrast,  this situation is alleviated in the MMV model.

Besides, given the same sparsity level (\emph{i.e}., the number of nonzero rows of $\mathbf{X}$ in the MMV model equals to the number of nonzero coefficients of $\mathbf{x}$ in the SMV model) and other same conditions, the MMV model can yield solutions with smaller errors than the SMV model \cite{Eldar2010average,Jin2013MMV}.

In practical use, the common sparsity constraint of the MMV model helps identify spectral peaks of MA in PPG spectra by using the spectral peaks in acceleration spectra. Since  MA components in PPG signals have many common frequencies with simultaneous acceleration signals, the common sparsity constraint encourages the frequency locations of MA in PPG spectra to be aligned well with some frequency locations in acceleration spectra. Consequently, one can accurately remove the spectral peaks of MA in the PPG spectra. This benefit is more desirable when the heartbeat frequency is very close to a frequency of MA.

\begin{figure}[t]
\centering
\includegraphics[width=9.5cm,height=10cm]{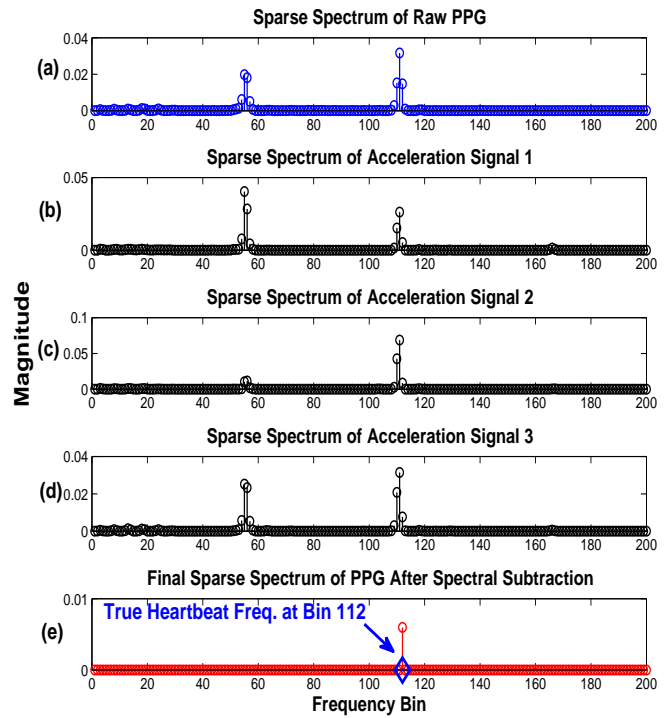}
\caption{Joint sparse signal recovery helps better identify and remove spectral peaks of MA in PPG spectra. (a)-(d) are sparse spectra of the PPG signal and three channels of acceleration signals in Fig.\ref{fig:signals}, respectively, which are calculated using the MMV model. (e) is the final sparse spectrum of the PPG signal after spectral subtraction. The true heartbeat frequency (estimated from the ECG signal) locates at the 112th frequency bin, which is accurately detected in (e). Note that in this example the HR frequency is very close to an MA frequency.}
\label{fig:MMVspectra}
\end{figure}

\begin{figure}[t]
\centering
\includegraphics[width=9.5cm,height=7cm]{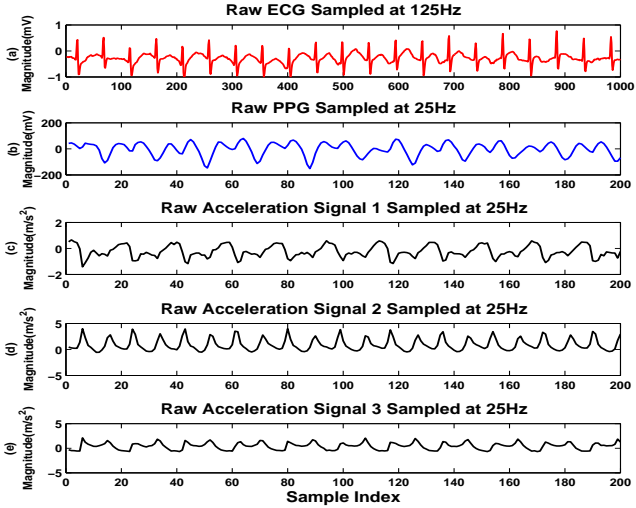}
\caption{Segments of simultaneously recorded raw signals. (a) A segment of the raw ECG signal sampled at 125 Hz (which provides ground-truth of HR). (b) A segment of the PPG signal sampled at 25 Hz. (c)-(e) are segments of the acceleration signals at three channels sampled at 25 Hz. }
\label{fig:signals}
\end{figure}

Fig.\ref{fig:MMVspectra} shows an example. Taking a PPG signal and  three simultaneous acceleration signals (shown in Fig.\ref{fig:signals}) to be the four measurement vectors in the MMV model, their spectra are jointly estimated, as shown in Fig.\ref{fig:MMVspectra}(a)-(d). Due to the common sparsity constraint on the estimated multichannel spectra, the frequency locations of MA in the PPG spectrum were exactly aligned with the frequency locations in the acceleration spectra. Thus, one can  subtract the MA spectral peaks in the acceleration spectra from the PPG spectrum (refer to the next subsection for details), yielding the cleansed spectrum shown in Fig.\ref{fig:MMVspectra}(e). We can see there was only one spectral peak remained, locating at the 112th frequency bin, which exactly corresponded to the heartbeat frequency (estimated from simultaneous ECG).

\begin{figure}[h]
\centering
\includegraphics[width=9.5cm,height=10cm]{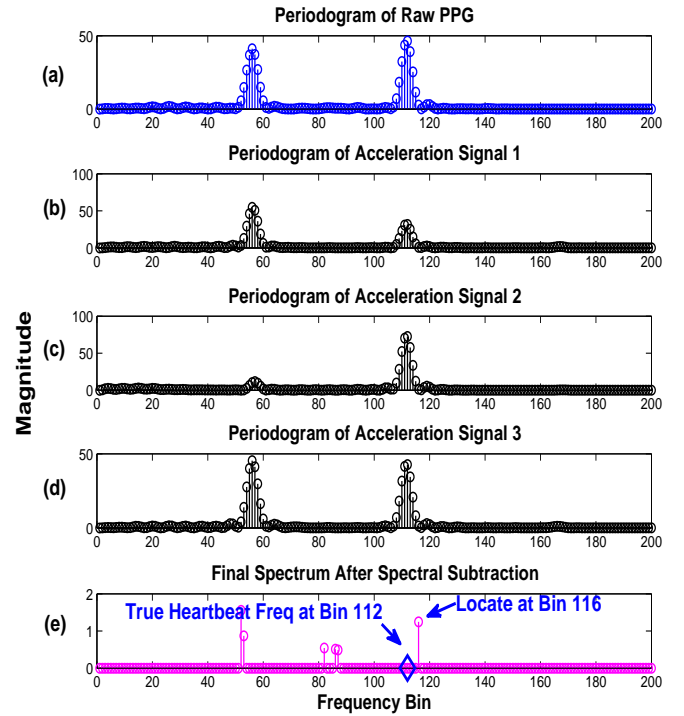}
\caption{Using Periodogram may result in large errors in HR estimation. (a)-(d) are spectra of the PPG signal and the three channels of acceleration signals in Fig.\ref{fig:signals}, respectively, which are calculated using the Periodogram separately. (e) is the final spectrum of the PPG signal after spectral subtraction as in Fig.\ref{fig:MMVspectra}. The true heartbeat frequency (located at the 112th frequency bin) is missed. The nearest spectral peak locates at the 116th frequency bin, indicating the error is about 5.9 BPM (each frequency bin corresponds to about 1.4 BPM). }
\label{fig:FFTspectra}
\end{figure}

Note that in this example the heartbeat frequency is close to the frequencies of MA. However, the proposed method correctly distinguishes the spectral peaks of MA from the spectral peak of heartbeat. This advantage cannot be gained by using conventional spectrum estimation algorithms such as Periodogram. Fig.\ref{fig:FFTspectra} shows the results by using Periodogram and the same spectral subtraction. Due to the low-resolution of Periodogram and the leakage effect \cite{TROIKA}, the spectral peak corresponding to the heartbeat is incorrectly removed during spectral subtraction. Instead, a false spectral peak is remained in the final spectrum, locating at the 116th frequency bin, indicating a large error of about 5.9 beat per minute (BPM).

Due to the advantages of the MMV model, finding the spectral peak corresponding to heartbeat becomes easier. In the following subsections, a simple spectral subtraction method will be first described, and then a spectral peak tracking method is proposed.

\subsection{Spectral Subtraction}
\label{subsec:spectral_subtraction}

Thanks to the advantages of the MMV model,  spectral subtraction is made simple (given one PPG spectrum and three acceleration spectra):
\begin{itemize}
  \item \textbf{Step 1}: For each frequency bin $f_i (i=1,\cdots,N)$,  choose the maximum value of spectral coefficients in acceleration spectra at $f_i$, denoted by $C_i$.

  \item \textbf{Step 2}: For each frequency bin $f_i (i=1,\cdots,N)$, subtract $C_i$ from the value of the spectral coefficient at $f_i$ in the PPG spectrum. Now we obtain a processed PPG spectrum. Denote by $p_{\mathrm{max}}$ the maximum value of all coefficients in the PPG spectrum.

  \item \textbf{Step 3}: Set to zero all spectral coefficients with values less than $p_{\mathrm{max}}/4$, yielding a cleansed PPG spectrum.
\end{itemize}

To ensure the spectral subtraction method effective, each of the PPG segment and acceleration  segments should be normalized to have the same variance (\emph{i.e.}, the same energy).

\subsection{Spectral Peak Tracking}
\label{subsec:selection}

As in TROIKA, spectral peak tracking is another key part. In this work, the proposed spectral peak tracking approach is simpler with less tuning parameters than the one in TROIKA. The approach is also based on the observation that HR values in two successive time windows are very close if the two time windows overlap largely.

The spectral peak tracking method consists of four stages, namely initialization, peak selection, peak verification, and peak discovery.

\subsubsection{Initialization}

In the initialization stage wearers are required to reduce hand motions as much as possible for several seconds, and HR is estimated by choosing the highest spectral peak in the PPG spectrum.

In the proposed approach, the kurtosis of the PPG spectrum from 0.8 Hz to 2.5 Hz (corresponding to 48 BPM to 150 BPM) is used to classify whether hand motions are reduced sufficiently. When hand motions occur, the kurtosis is small; otherwise, it is very large. Thus, in the proposed approach, if the kurtosis is larger than 10, the current time window is determined to be in the initialization stage, and  the highest spectral peak in the PPG spectrum from 0.8 Hz to 2.5 Hz is chosen; in the next time window the approach enters the next stage. If the kurtosis is smaller than 10, the current time window is not in the initialization stage, and thus no HR estimate is output; in the next time window, the approach still checks whether it is in the initialization stage.

\subsubsection{Peak Selection}
\label{subsubsec:peakSelection}

\begin{figure}[t]
\centering
\includegraphics[width=9cm,height=11cm]{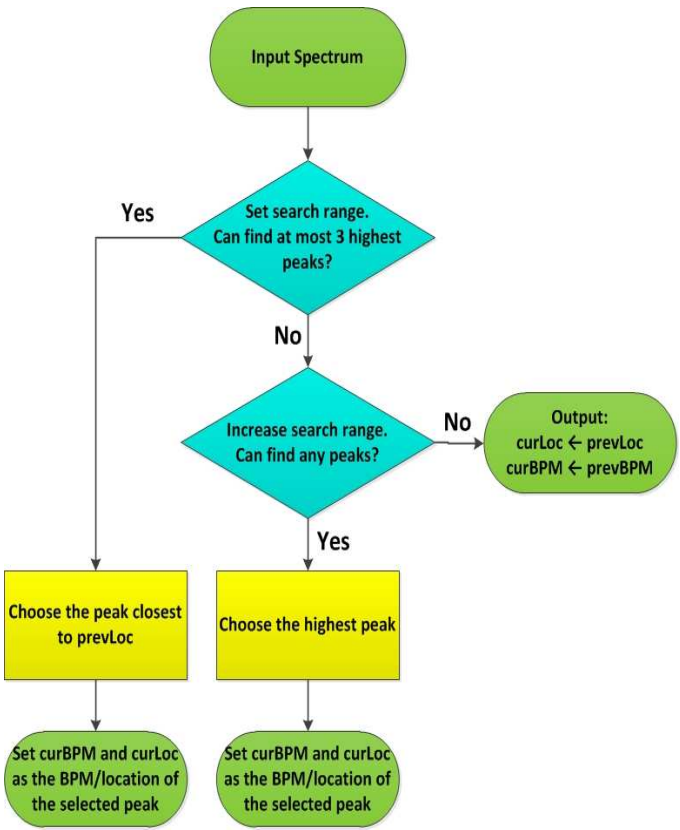}
\caption{Flowchart of the peak selection procedure.}
\label{fig:peakSelection}
\end{figure}

In this stage, the goal is to choose a  peak in the PPG spectrum with the knowledge of  estimated HR values in previous time windows. The flowchart is given in Fig.\ref{fig:peakSelection}.

First,  a search range is set, denoted by $R_1$. This search range is centered at the location of the previously estimated spectral peak. In particular, denote by $\mathrm{prevLoc}$ the frequency bin corresponding to the previously estimated HR, and denote by $\mathrm{prevBPM}$ the corresponding BPM value. Then $R_1 \triangleq [\mathrm{prevLoc} - \Delta_1, \; \mathrm{prevLoc} + \Delta_1]$, where $\Delta_1$ is a positive integer.

Next, in the search range $R_1$ we seek at most 3 spectral peaks. If find any, then choose the one \textbf{closest} to $\mathrm{prevLoc}$. If not, then set another search range $R_2 \triangleq  [\mathrm{prevLoc} - \Delta_2, \; \mathrm{prevLoc} + \Delta_2]$ with $\Delta_2 > \Delta_1$. If in this range we still cannot find any peak, then output $\mathrm{prevLoc}$ and  $\mathrm{prevBPM}$. If find any, then select the one with the \textbf{highest} value.

When selecting the peak closest to $\mathrm{prevLoc}$, one may face a situation that  two peaks have equal distance to $\mathrm{prevLoc}$. (One has lower frequency and another has higher frequency.) To decide which peak should be selected, it is needed to predict whether current HR value will increase or decrease compared to the previously estimated HR value. This can be efficiently done by performing the Smoother algorithm in \cite{Eilers2003perfect} on the spectral locations of $H$ previously estimated HR values, where $H \geq 10$. The smoothing parameter in the algorithm should be set to a small value such that we can predict local change of HR values.

Let $\mathrm{curLoc}$ and $\mathrm{curBPM}$ denote the frequency bin and associated BPM of the selected spectral peak, respectively.

\begin{figure}[t]
\centering
\includegraphics[width=9cm,height=11cm]{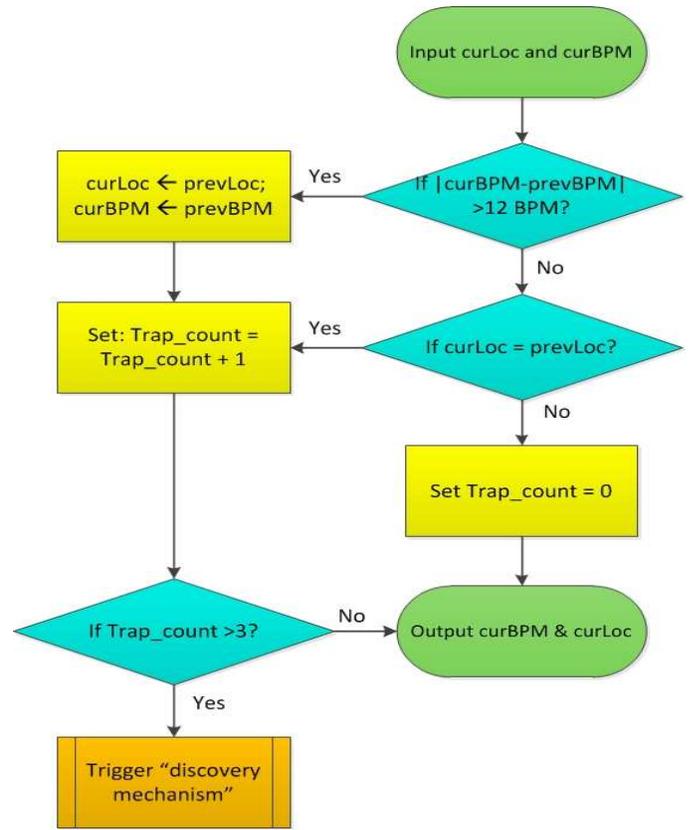}
\caption{Flowchart of the peak verification procedure.}
\label{fig:peakVerification}
\end{figure}

\begin{figure}[t]
\centering
\includegraphics[width=9cm,height=11cm]{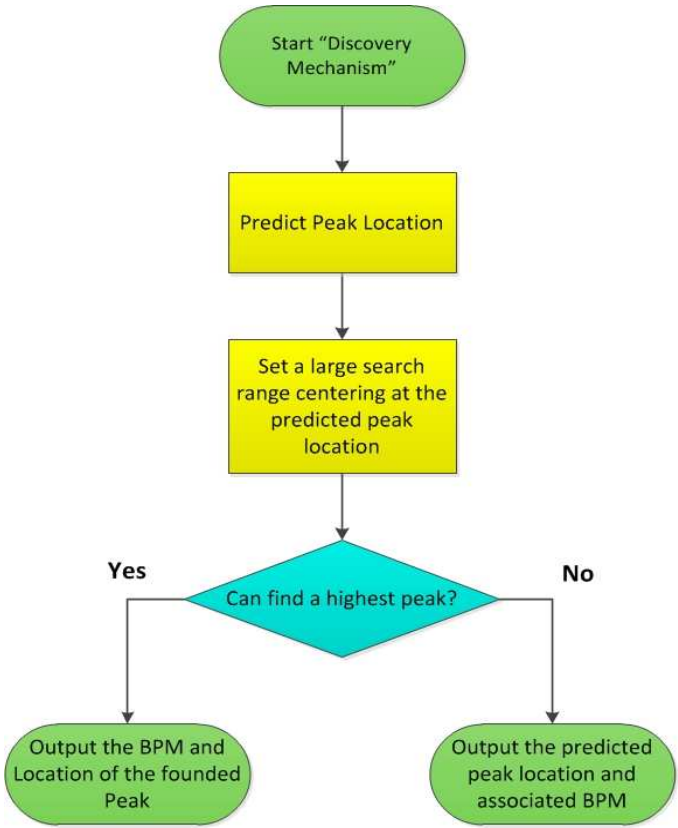}
\caption{Flowchart of the peak discovery procedure.}
\label{fig:peakDiscovery}
\end{figure}

\subsubsection{Peak Verification}

The previous stage sometimes can wrongly select a spectral peak associated with MA. Thus the verification stage is necessary. The flowchart is given in Fig.\ref{fig:peakVerification}. The verification method is based on an observation that the change of BPM values in two successive time windows rarely exceeds 12 BPM. If the difference between $\mathrm{curBPM}$ and $\mathrm{prevBPM}$ is larger than 12 BPM, then it is highly possible that the selected spectral peak is incorrect. So,  $\mathrm{curLoc}$ and $\mathrm{curBPM}$ are reset to $\mathrm{prevLoc}$ and $\mathrm{prevBPM}$, respectively.

However, when $\mathrm{curLoc}=\mathrm{prevLoc}$ happens in multiple successive time windows, this indicates that the target spectral peak is lost. Thus, a discovery procedure is triggered.

\subsubsection{Peak Discovery}

Fig.\ref{fig:peakDiscovery} shows the flowchart of peak discovery. The Smoother algorithm is performed on the spectral locations of $K$ previously estimated HR values to predict the spectral location of heartbeat in current time window. Different to the prediction in Peak Selection Stage, here the goal is to predict a macro-trend of HR evolution. So $K$ is much larger than $H$, and the smoothing parameter is set to a large value. Since there is large uncertainty on the exact spectral location of HR,  a larger search range is set than before. In particular, the search range is $R_3 \triangleq [\mathrm{predictLoc} - \Delta_3, \; \mathrm{predictLoc} + \Delta_3]$, where $\mathrm{predictLoc}$ is the predicted location, and $\Delta_3$ is a large integer.

\section{Experimental Results}
\label{sec:experiments}

\subsection{Datasets}

The JOSS algorithm was evaluated on the 12 PPG datasets initially used in \cite{TROIKA} \footnote{The datasets are also the training sets of 2015 IEEE Signal Processing Cup: \url{http://icassp2015.org/signal-processing-cup-2015/}.}. Each dataset contains a channel of PPG, three channels of acceleration signals, and a channel of ECG, all recorded simultaneously from 12 healthy male subjects with age ranging from 18 to 35. In each dataset, the PPG signal was recorded from the wrist (dorsal) using a  pulse oximeter   with green LED (wavelength: 515nm). The three-channel acceleration signal was also recorded from  the wrist using a three-axis accelerometer. Both the pulse oximeter and the accelerometer were embedded in a wristband. The ECG signal was recorded from the chest using wet ECG electrodes, from which the ground-truth of HR was calculated to evaluate algorithm performance. All signals were sent to a nearby computer via Bluetooth.

During data recording the subjects walked or ran on a treadmill with the following speeds in order: the speed of 1-2 km/hour for 0.5 minute, the speed of 6-8 km/hour for 1 minute, the speed of 12-15 km/hour for 1 minute, the speed of 6-8 km/hour for 1 minutes, the speed of 12-15 km/hour for 1 minute, and the speed of 1-2 km/hour for 0.5 minute.

All signals were initially sampled at 125 Hz. But in the present experiments the PPG signals and the acceleration signals were downsampled  to 25 Hz. Some segments of these signals are shown in Fig.\ref{fig:signals}.

\subsection{Experimental Settings}

As in \cite{TROIKA},  a time window of 8 seconds was used to slide the simultaneous PPG signal and acceleration signals, with a step of 2 seconds.  HR was estimated in each time window.  Before performing JOSS, all raw signals were bandpass filtered from 0.4 Hz to 4 Hz using the 2nd order Butterworth filter.

The Regularized M-FOCUSS algorithm \cite{Cotter2005} was used to estimate the solution matrix of the MMV model, with the parameter $p=0.8$, the regularization parameter $\lambda = 10^{-10}$, and the spectrum grid number $N = 1024$. Its maximum iteration number was set to 4. Note that the TROIKA algorithm also used the M-FOCUSS algorithm to estimate sparse spectrum of PPG signals. However, in TROIKA, the M-FOCUSS algorithm was used to estimate the solution of the SMV model \footnote{Since the SMV model is a special case of the MMV model (\ref{eq:MMV}) when $L=1$, almost all MMV-model-based algorithms can work on the SMV model. But in this case, the benefits of exploiting multiple measurement vectors do not exhibit.}.

In the part of spectral peak tracking,  $\Delta_1 = 15$, $\Delta_2=25$, and $\Delta_3=30$. The smoothing parameter used in the Peak Selection Stage was set to 5 and $H=10$, while in the Peak Discovery Stage the smoothing parameter was set to 20 and $K=30$.

\subsection{Performance Measurement}
\label{subsec:performance}

The ground-truth HR was calculated from the original raw ECG signals (sampled at 125 Hz)  by manually picking the R-peaks one by one in each time widnow. No R-peak detection algorithm was used, in order to avoid any possible detection errors. These ground-truth HR values are now available in the PPG datasets \cite{TROIKA} for performance evaluation.

The  measurement indexes in \cite{TROIKA} were used. One was the average absolute  error (in BPM), defined as
\begin{eqnarray}
\mathrm{Error1} = \frac{1}{W} \sum_{i=1}^W \big|\mathrm{BPM}_\mathrm{est}(i) -  \mathrm{BPM}_\mathrm{true}(i) \big|
\end{eqnarray}
where  $\mathrm{BPM}_\mathrm{true}(i)$ denotes the ground-truth of HR in the $i$-th time window, $\mathrm{BPM}_\mathrm{est}(i)$ denotes the estimated HR, and $W$ is the total number of time windows.

The second was the average absolute error percentage, defined as
\begin{eqnarray}
\mathrm{Error2} = \frac{1}{W} \sum_{i=1}^W \frac{\big|\mathrm{BPM}_\mathrm{est}(i) -  \mathrm{BPM}_\mathrm{true}(i) \big|}{\mathrm{BPM}_\mathrm{true}(i) }.
\end{eqnarray}

The third index was the Bland-Altman plot \cite{martin1986statistical}. The Limit of Agreement (LOA) was used, which is defined as $[\mu-1.96\sigma,\mu+1.96\sigma]$, where $\mu$ is the average difference between  each estimate and the associated ground-truth against their average, and $\sigma$ is the standard deviation.

Pearson correlation between  ground-truth values and  estimates was also adopted.

\begin{table*}[t]
\renewcommand{\arraystretch}{1.2}
\label{table:err1}
\centering
\begin{threeparttable}
\caption{Comparison of JOSS and TROIKA in terms of average absolute errors (Error1) on the 12 datasets sampled at 25 Hz. The unit is BPM. SD denotes standard deviation. }
\begin{tabular}{l|c|c|c|c|c|c|c|c|c|c|c|c|c}
\toprule
       &  Set 1      & Set 2   &  Set 3   &  Set 4   & Set 5  & Set 6  & Set 7  & Set 8  &Set 9  &Set 10 &Set 11 & Set 12 & \textbf{Average}\\
\midrule
JOSS       &   1.33  &  1.75   &  1.47   &  1.48    & 0.69   &  1.32 & 0.71 & 0.56 & 0.49 & 3.81  & 0.78  & 1.04 & \textbf{1.28 (SD=2.61)}\\
\hline
TROIKA       &   2.87   &  2.75   &  1.91   &  2.25    & 1.69   &   3.16 & 1.72 & 1.83 & 1.58 & 4.00  & 1.96  & 3.33 & \textbf{2.42 (SD=2.47)}\\
\hline
\bottomrule
\end{tabular}
\end{threeparttable}
\end{table*}

\begin{table*}[t]
\renewcommand{\arraystretch}{1.2}
\label{table:err2}
\centering
\begin{threeparttable}
\caption{Comparison of JOSS and  TROIKA  in terms of average absolute error percentage (Error2) on the 12 datasets sampled at 25 Hz. SD denotes standard deviation.}
\begin{tabular}{l|c|c|c|c|c|c|c|c|c|c|c|c|c}
\toprule
       &  Set 1      & Set 2   &  Set 3   &  Set 4   & Set 5  & Set 6  & Set 7  & Set 8  &Set 9  &Set 10 &Set 11 & Set 12 & \textbf{Average}\\
\midrule
JOSS      &  1.19\%   &  1.66\%   &  1.27\%  & 1.41\%    &  0.51\%   &  1.09\% & 0.54\% &  0.47\% &  0.41\% &  2.43\% & 0.51\% & 0.81\% & \textbf{1.01\% (SD=2.29\%)}\\
\hline
TROIKA      &  2.18\%   &  2.37\%   &  1.50\%  & 2.00\%    &  1.22\%   &  2.51\% & 1.27\% &  1.47\% &  1.28\% &  2.49\% & 1.29\% & 2.30\% & \textbf{1.82\% (SD=2.07\%)}\\
\hline
\bottomrule
\end{tabular}
\end{threeparttable}
\end{table*}

\begin{figure}[t]
\centering
\includegraphics[width=9cm,height=5.5cm]{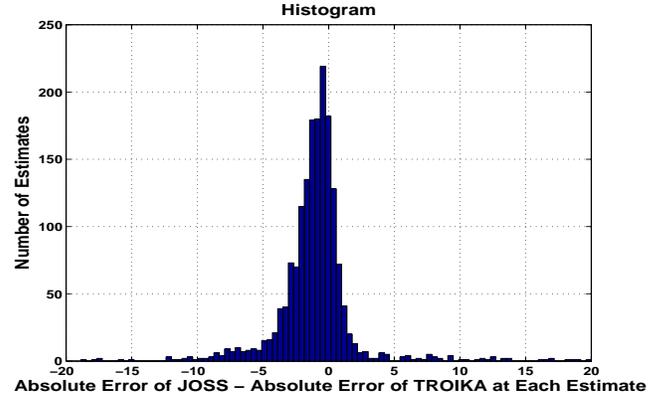}
\caption{Histogram of the difference between the absolute error of JOSS and the absolute error of TROIKA at each estimate over the 12 datasets. Based on the $t$-test, the absolute estimation error of JOSS was significantly smaller than that of TROIKA at the significant level $a=0.01$, and the $p$ value was $6.3 \times 10^{-39}$. }
\label{fig:err}
\end{figure}

\subsection{Results}

Table~I and Table~II list the average absolute error (Error1) and the average absolute error percentage (Error2) of the proposed JOSS algorithm on all 12 datasets, respectively \footnote{For several datasets, the initial recordings contain strong MA, probably due to  device adjustment after the recording system turned on. For these recording segments, both algorithms did not output HR estimates. Thus the algorithms' performance was not evaluated on these segments. These excluded segments were the first 12 seconds of Set 2, the first 8 seconds of Set 3, the first 2 seconds of Set 4, the first 2 seconds of Set 8, the first 6 seconds of Set 10, and the first 2 seconds of Set 11.}. To show its superiority, it was compared with the recently proposed TROIKA algorithm. For direct comparison, TROIKA was also performed on the PPG and acceleration data sampled at 25 Hz. The results in Table~I and Table~II show that JOSS had  better performance than TROIKA. Averaged across the 12 datasets, the absolute estimation error (Error1) of JOSS was $1.28 \pm 2.61$ BPM (mean $\pm$ standard deviation), and the error percentage (Error2) was $1.01\% \pm 2.29\%$. In contrast, the Error1 of TROIKA was $2.42 \pm 2.47$ BPM, and the Error2 was $1.82\% \pm 2.07\%$.

To better compare the performance of JOSS with TROIKA, Fig.\ref{fig:err} plots the histogram of the difference between the absolute error of JOSS and the absolute error of TROIKA at each estimate over the 12 datasets, \emph{i.e.}, the histogram of $\alpha(i)-\beta(i)$, where $\alpha(i)$ indicates the absolute estimation error of JOSS at the $i$th heart rate estimate, and $\beta(i)$ indicates that of TROIKA at the $i$th estimate. Based on the $t$-test, the absolute estimation error of JOSS was significantly smaller than that of TROIKA at the significant level $a=0.01$, and the $p$ value was $6.3 \times 10^{-39}$.

\begin{figure}[t]
\centering
\includegraphics[width=9cm,height=5.5cm]{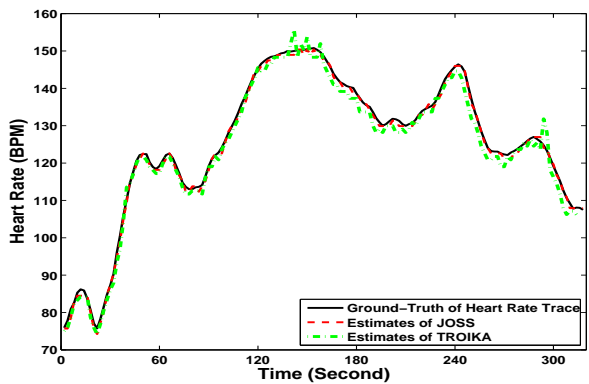}
\caption{Estimation results on Dataset 8. The estimated HR trace by JOSS was almost the same as the ground-truth of HR trace, while TROIKA  got errors sometimes.}
\label{fig:data8}
\end{figure}

Fig.\ref{fig:data8} shows the estimated HR traces of both JOSS and TROIKA on Dataset 8 (randomly chosen). JOSS had better performance than TROIKA; its estimated HR trace was almost the same as the ground-truth of HR trace, while TROIKA  sometimes got errors.

\begin{figure}[t]
\centering
\includegraphics[width=9cm,height=5.5cm]{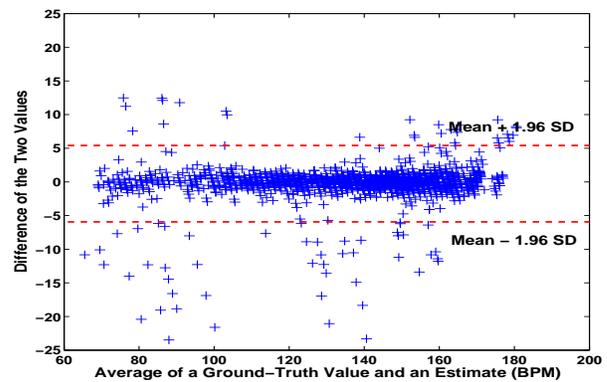}
\caption{The Bland-Altman plot of the estimation results on the 12 datasets. The LOA was $[-5.94, 5.41]$ BPM.}
\label{fig:BAplot}
\end{figure}

\begin{figure}[t]
\centering
\includegraphics[width=9cm,height=5.5cm]{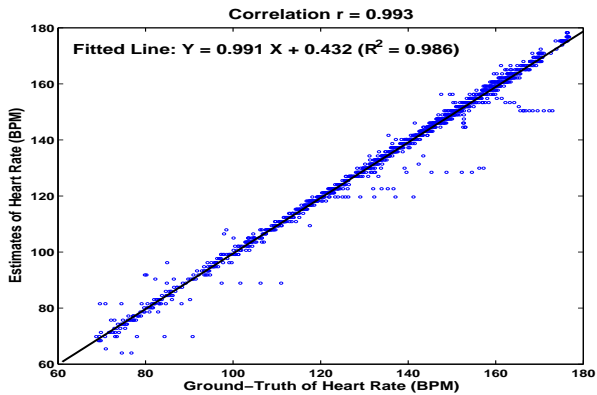}
\caption{Scatter plot between the ground-truth heart rate values and the associated estimates over the 12 datasets. The fitted line was $y=0.991 x + 0.432$, where $x$ indicates the ground-truth heart rate value, and $y$ indicates the associated estimate. The $R^2$ value, a measure of goodness of fit, was 0.986. The Pearson correlation was 0.993.}
\label{fig:corrPlot}
\end{figure}

Fig.\ref{fig:BAplot} gives the Bland-Altman plot. The LOA was $[-5.94, 5.41]$ BPM. The Scatter Plot between the ground-truth heart rate values and the associated estimates over the 12 datasets is given in Fig.\ref{fig:corrPlot}, which shows the fitted line was $y=0.991 x + 0.432$, where $x$ indicates the ground-truth heart rate value, and $y$ indicates the associated estimate. The Pearson coefficient was 0.993.

\section{Discussions}
\label{sec:discussion}

\subsection{On the Experiments}

In the experiments the subjects had yellow skin-color and the LED light was green. Skin-color and LED light are known factors that affect characteristic of PPG signals, and thus affect algorithm performance. Evaluating the proposed algorithm's performance at other skin-color and LED light  will be the future work.

Studies have  shown that the PPG pulse rate variability is a surrogate of heart rate variability when subjects are at rest \cite{gil2010photoplethysmography}. However, when body movements occur, the correlation between PPG pulse-to-pulse intervals and ECG beat-to-beat intervals is not very high, and the correlation varies largely at different anatomical measurement sites \cite{maeda2011relationship}. Therefore, it is more reasonable to compare the ground-truth HR and the estimated HR in an `average' level, instead of comparing each single cardiac cycle period. In the present experiments, the ground-truth HR was calculated from ECG in the time window of 8 seconds by the formula $60 C/D$ (in BPM), where $C$ was the number of cardiac cycles in the time window and $D$ was the duration (in seconds) of these cycles \cite{TROIKA}. And the estimated HR  was calculated from the spectrum of PPG in the same time window. These calculation methods actually obtain averaged heartbeat periods in each time window, which is relatively resistant to the beat-to-beat interval variability.

\subsection{Advantages and Possible Improvement Approaches of JOSS}

As shown in the experiments, JOSS works well at low sampling rates. This is a desirable advantage over algorithms that only work well at high sampling rates, since a low sampling rate means low energy consumption in data acquisition and in wireless transmission, which can largely extend battery life in wearable devices.

JOSS exploits a common sparsity structure in the spectra of PPG signals and simultaneous acceleration signals by using the MMV model in sparse signal recovery. The MMV model is known to be superior to the basic sparse signal recovery model used in TROIKA. It is worth noting that there are many other structures in the spectra of PPG and acceleration signals which can be exploited. Thus, more advanced sparse signal recovery models may be used to further improve spectrum estimation performance and HR estimation performance, such as the model exploiting magnitude correlation among spectra \cite{Zhang2011IEEE,Zhang2014IEEE}.

Gridless joint spectral compressed sensing \cite{Chi2014jointCS,Duarte2013spectral} is another promising framework to reduce HR estimation errors. When HR frequency does not locate at a frequency grid,  estimation errors are unavoidable, and the PPG spectrum calculated from the MMV model may be less-sparse. This may result in some difficulties for conventional sparse signal recovery algorithms. But gridless joint spectral compressed sensing can potentially solve this issue.

There are other possible approaches to improve HR estimation. One approach is applying smoothing techniques to estimated HR traces, such as the median filtering. However, most effective smoothing algorithms perform offline. Thus this technique may be an effective approach when strict real-time is not required.

Although JOSS does not require an extra noise-removal modular, using a noise-removal modular can improve its robustness to MA. For example, JOSS can also adopt the signal decomposition modular used in TROIKA to partially remove MA before joint sparse spectrum reconstruction. Of course, this increases  processing time, power consumption, and circuit design complexity if implemented in VLSI or FPGA.

%

\section{Conclusion}

In this work a PPG-based heart rate monitoring method was proposed for fitness tracking via smart-watches or other wearable devices. The method uses the multiple measurement vector model in sparse signal recovery to jointly estimate  sparse spectra of PPG signals and simultaneous acceleration signals. Due to the common sparsity constraint on the spectral coefficients, identifying and removing spectral peaks of MA in PPG spectra is easier. Thus, it does not need an extra signal processing stage to remove MA as in other algorithms. This largely simplifies the whole algorithm. Besides, it works well for data sampled at low sampling rates, thus saving energy consumption in data acquisition and wireless transmission. Therefore, the proposed method has potential to be implemented in VLSI or FPGA in wearable devices.

\section*{Acknowledgement}
This work was supported in part by Samsung Research America--Dallas. But any opinions, findings, and conclusions expressed in this work are those of the author and do not necessarily reflect the views of the funding organization.

\bibliographystyle{IEEEtran}

\bibliography{bib_PPG,bib_CS,bib_Spectrum,bib_SBL,bib_Zhilin}

\begin{biography}[{\includegraphics[width=1.0in,height=1.25in,clip,keepaspectratio]{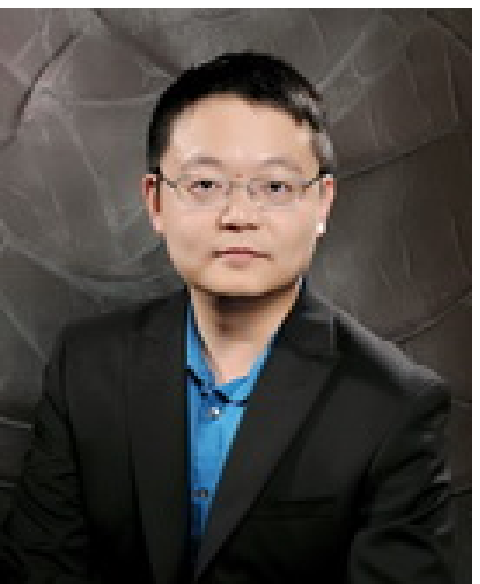}}]{Zhilin Zhang} (SM'15) received the Ph.D. degree in electrical engineering from University of California at San Diego, La Jolla, CA, USA, in 2012. He is currently a staff research engineer and manager with the Emerging Technology Lab in Samsung Research America -- Dallas, Texas, USA, and an adjunct professor with School of Computer Science and Engineering, University of Electronic Science and Technology of China, Sichuan, China.

His research interests include sparse signal recovery/compressed sensing, statistical signal processing, biomedical signal processing, time series analysis, machine learning, and their applications in wearable healthcare, smart-home, and finance. He has published about 40 papers in peer-reviewed journals and conferences.

He is a member of the Bio-Imaging and Signal Processing Technical Committee of the IEEE Signal Processing Society, a Technical Program Committee Member of a number of international conferences, and a main organizer of the 2015 IEEE Signal Processing Cup. He is currently an Associate Editor of IEEE JOURNAL OF TRANSLATIONAL ENGINEERING IN HEALTH AND MEDICINE.

He received Excellent Master Thesis Award in 2005, the Second Prize in College Student Entrepreneur Competition (for a fetal heart rate monitoring product) in 2005, and Samsung Achievement Awards in 2013 and 2014. Two of his papers published in IEEE TRANSACTIONS ON BIOMEDICAL ENGINEERING were ranked as the \emph{Most Cited Articles Published in 2013 and 2014} in the journal.

\end{biography}

\end{document}